\documentstyle[11pt,epsfig]{article}
\begin{document}
\input amssym.def 
\input amssym
\hfuzz=5.0pt
%
%
%
%
\def\vec#1{\mathchoice{\mbox{\boldmath$\displaystyle\bf#1$}}
{\mbox{\boldmath$\textstyle\bf#1$}}
{\mbox{\boldmath$\scriptstyle\bf#1$}}
{\mbox{\boldmath$\scriptscriptstyle\bf#1$}}}
\def\mbf#1{{\mathchoice {\hbox{$\rm\textstyle #1$}}
{\hbox{$\rm\textstyle #1$}} {\hbox{$\rm\scriptstyle #1$}}
{\hbox{$\rm\scriptscriptstyle #1$}}}}
\def\operatorname#1{{\mathchoice{\rm #1}{\rm #1}{\rm #1}{\rm #1}}}
\chardef\ii="10
\def\widehat{\mathaccent"0362 }
\def\widetilde{\mathaccent"0365 }
\def\vphi{\varphi}
\def\vrho{\varrho}
\def\vtheta{\vartheta}
\def\ih{{\i\over\hbar}}
\def\CD{{\cal D}}
\def\CL{{\cal L}}
\def\CP{{\cal P}}
\def\CV{{\cal V}}
\def\half{{1\over2}}
\def\bhalf{\hbox{$\half$}}
\def\viert{{1\over4}}
\def\bviert{\hbox{$\viert$}}
\def\dfrac#1#2{\frac{\displaystyle #1}{\displaystyle #2}}
\def\pathint#1{\int\limits_{#1(t')=#1'}^{#1(t'')=#1''}\CD #1(t)}
\def\hbarm{{\dfrac{\hbar^2}{2m}}}
\def\pmb#1{\setbox0=\hbox{#1}
    \kern-.025em\copy0\kern-\wd0
    \kern.05em\copy0\kern-\wd0
    \kern-.025em\raise.0433em\box0}
\def\bbbr{{\rm I\!R}}                                
\def\bbbn{{\rm I\!N}}                                
\def\bbbz{{\mathchoice {\hbox{$\sf\textstyle Z\kern-0.4em Z$}}
{\hbox{$\sf\textstyle Z\kern-0.4em Z$}}
{\hbox{$\sf\scriptstyle Z\kern-0.3em Z$}}
{\hbox{$\sf\scriptscriptstyle Z\kern-0.2em Z$}}}}    
\def\Cl{\operatorname{Cl}} 
\def\dt{\d t}
\def\d{\operatorname{d}}
\def\e{\operatorname{e}}
\def\i{\operatorname{i}}
 
\begin{titlepage}
\centerline{\normalsize DESY 98--100 \hfill ISSN 0418 - 9833}
\centerline{\normalsize August 1998\hfill}
\centerline{\normalsize quant-ph/9808016\hfill}
\vskip.3in
\message{TITLE:}
\message{PATH INTEGRALS WITH KINETIC COUPLING POTENTIALS}
\begin{center}
{\Large PATH INTEGRALS WITH KINETIC 
 COUPLING POTENTIALS}
\end{center}
\vskip.3in
\begin{center}
{\Large Christian Grosche}
\vskip.2in
{\normalsize\em II.\,Institut f\"ur Theoretische Physik}
\vskip.05in
{\normalsize\em Universit\"at Hamburg, Luruper Chaussee 149}
\vskip.05in
{\normalsize\em 22761 Hamburg, Germany}
\end{center}
\vfill
\begin{center}
{ABSTRACT}
\end{center}
\smallskip
\noindent
{\small
Path integral solutions with kinetic coupling potentials $\propto p_1p_2$
are evaluated. As examples I give a Morse oscillator, i.e., a model in 
molecular physics, and the double pendulum in the harmonic approximation.
The former is solved by some well-known path integral techniques, whereas the 
latter by an affine transformation.}

\end{titlepage}
 
 
\normalsize
\section{Introduction}
Quantum mechanics is about physics on the atomic level, i.e., wave-functions
and energy levels of atoms and molecules. The simplest quantum mechanical
system is the harmonic oscillator with the characteristic $\hbar(n+\half)$
spectrum \cite{FEYb}--\cite{SCHUHd} which was also the first system solved by 
means of the path integral. The important Coulomb system, i.e., the hydrogen 
atom, was only solved later on by Duru and Kleinert \cite{DKa,KLEo}.
In this Paper I want to demonstrate the technique of path integration for 
some particular models with kinetic coupling. The first example consists 
of the potential force of two molecules, where the anharmonic interaction
force is described by  the Morse potential with an additional kinetic
coupling $\propto p_1p_2$, respectively $\propto\dot x_1\dot x_2$, the
indices $1$ and $2$ referring to particle one and two. The second example is the
double pendulum, a system which is coupled by a $\propto\dot\vphi_1\dot\vphi_2$
term. In the harmonic approximation the double pendulum is described by
two coupled oscillators. In fact, these two examples are not too difficult to solve,
but they demonstrate in a nice way the applicability of path integration techniques
for systems which are coupled by means of the kinetic energy. Such problems 
seem not to have been studied by path integrals yet, whereas systems which are 
coupled through only a coordinate dependence, i.e.~$\propto x_1x_2$, can be found 
in e.g.~\cite{GRSh}.

\section{Morse Oscillator with Kinetic Coupling}
The first example is a model in molecular physics, which takes into account the 
anharmonicity of the inter-molecular potential (e.g., H$_2$O, H$_2$S, C$_2$H$_2$), 
by means of the Morse potential \cite{MIRO}--\cite{FANCF}
\begin{equation}
  V(x_1-x_2)=\lambda\Big(1-\alpha\,\e^{-\beta(x_1-x_2-r_0)}\Big)^2
  \enspace,
\end{equation}
where $x_1,x_2$ denote the position of the first and second particle one with 
masses $m_1,m_2$, $r_0$ the equilibrium position, and $\alpha,\beta,\lambda$ 
are positive parameters. $\hbar$ denotes Planck's constant divided by $2\pi$.
The Hamiltonian of this system for the two particles has the form (with the 
momentum operators $P_{1,2}=(\hbar/\i)\partial_{x_{1,2}}$)
\begin{equation}
  H=\dfrac{1}{2m_1}P_1^2+\dfrac{1}{2m_2}P_2^2+\kappa P_1P_2
  +\lambda\Big(1-\alpha\,\e^{-\beta(x_1-x_2-r_0)}\Big)^2
  \enspace.
\end{equation}
The corresponding Hamiltonian path integral for the time evolution from
$x_{1,2}(t')$ to $x_{1,2}(t'')$ in the time interval $T=t''-t'$ can thus be 
written as \cite{GRSh}
\begin{eqnarray}       & &\!\!\!\!\!\!\!\!
  K(x_1'',x_1',x_2'',x_2';T)
  =\pathint{x_1}\int\CD P_1(t)\pathint{x_2}\int\CD P_2(t)\qquad\qquad
         \nonumber\\   & &\!\!\!\!\!\!\!\!\qquad\times
  \exp\Bigg\{\ih\int_{t'}^{t''}\Bigg[
  P_1\dot x_1+P_2\dot x_2-\dfrac{1}{2m_1}P_1^2-\dfrac{1}{2m_2}P_2^2
         \nonumber\\   & &\!\!\!\!\!\!\!\!
  \qquad\qquad\qquad\qquad \qquad\qquad\qquad\qquad
  -\kappa P_1P_2-\lambda\Big(1-\alpha\,\e^{-\beta(x_1-x_2-r_0)}\Big)^2
  \Bigg]\dt\Bigg\}\enspace. 
\end{eqnarray}
We can switch to center-of-mass $X$ and relative coordinates $x_r$ by means of
\begin{equation}
\left.\begin{array}{lll} 
  X=\mu_1 x_1+\mu_2x_2\enspace, &\qquad &x_r=x_1-x_2-r_0        \enspace,
  \\[3mm]   
  P=P_1+P_2           \enspace, &\qquad &P_r=\mu_2 P_1-\mu_1 P_2\enspace,
\end{array}\qquad\right\}
\end{equation}
where $(i=1,2)$  $\mu_i=m_i/M$, $(\mu_1+\mu_2=1)$, $M=m_1+m_2$ is the total mass, 
and $\mu=m_1m_2/M$ is the reduced mass. Vice versa we have
\begin{equation}
x_1=X+\dfrac{m_2}{M}x_r\enspace,\quad
x_2=X-\dfrac{m_1}{M}x_r\enspace,\quad
P_1=P_r+\mu_1P\enspace,\quad
P_2=\mu_2P-P_r\enspace.
\end{equation}
The Hamiltonian path integral in center-of-mass and relative coordinates 
is therefore given by
\begin{eqnarray}       & &\!\!\!\!\!\!\!\!
  K(X'',X',x_r'',x_r';T)
  =\pathint{X}\int\CD P(t)\pathint{x_r}\int\CD P_r(t)\qquad\qquad
         \nonumber\\   & &\!\!\!\!\!\!\!\!\qquad\times
  \exp\Bigg\{\ih\int_{t'}^{t''}\Bigg[P\dot X+P_r\dot x_r
  -\bigg(\dfrac{1}{2M}+\kappa\mu_1\mu_2\bigg)P^2
  -\bigg(\dfrac{1}{2\mu}-\kappa\bigg)P_r^2
         \nonumber\\   & &\!\!\!\!\!\!\!\!\qquad\qquad\qquad\qquad\qquad\qquad
  -\kappa(\mu_2-\mu_1)PP_r
  -\lambda\Big(1-\alpha\,\e^{-\beta x_r}\Big)^2\Bigg]\dt\Bigg\}
  \enspace. 
\end{eqnarray}
We can perform the momenta integrations to obtain the Lagrangian
path integral for the coupled system
\begin{eqnarray}       & &\!\!\!\!\!\!\!\!
  K(X'',X',x_r'',x_r';T)=\pathint{X}\pathint{x_r}
         \nonumber\\   & &\!\!\!\!\!\!\!\!\qquad\times
  \exp\Bigg\{\ih\int_{t'}^{t''}\Bigg[
  \dfrac{d}{4D}\dot X^2+\dfrac{a}{4D}\dot x_r^2-2\dfrac{b}{4D}\dot x_r\dot X
  -\lambda\Big(1-\alpha\,\e^{-\beta x_r}\Big)^2\Bigg]\dt\Bigg\}\enspace,
\label{pathXxr}
\end{eqnarray}
and I have used the abbreviations
\begin{equation}
a=\dfrac{1}{2M}+\kappa\mu_2\mu_2\enspace,\quad
d=\dfrac{1}{2\mu}-\kappa\enspace,\quad
b=\dfrac{\kappa}{2}(\mu_2-\mu_1)\enspace,
\end{equation}
and $D=ad-b^2=\bviert\big(1/m_1m_2-\kappa^2\big)$. The limiting case $\kappa=0$
is easily recovered. The lattice formulation of the path integral (\ref{pathXxr}) 
is given in the usual way by
\begin{eqnarray}       & &\!\!\!\!\!\!\!\!
  K(X'',X',x_r'',x_r';T)
  =\lim_{N\to\infty}\bigg(\dfrac{1}{4\sqrt{D}\,\pi\i\epsilon\hbar}\bigg)^N
  \prod_{j=1}^{N-1}\int_{\bbbr}\d X_j\int _{\bbbr} \d x_{r_j}
         \nonumber\\   & &\!\!\!\!\!\!\!\!\qquad\times
  \exp\Bigg\{\ih\sum_{j=1}^N\Bigg[
  \dfrac{d}{4D}(\Delta X_j)^2+\dfrac{a}{4D}(\Delta x_{r_j})^2
  -\dfrac{b}{2D} \Delta X_j \Delta x_{r_j}
  -\lambda\Big(1-\alpha\,\e^{-\beta x_{r_j}}\Big)^2\Bigg]\Bigg\}\enspace.
         \nonumber\\   & &
\end{eqnarray}
Here we have taken as usual $\epsilon=T/N=(t''-t')/N$ in the time slicing,
$X_j=X(t_j), x_{r_j}=x_r(t_j)$, $(j=1,\dots,N)$, $\Delta X_j=X_j-X_{j-1},
\Delta x_{r_j}=x_{r_j}-x_{r_{j-1}}$, as $N\to\infty,\epsilon\to0$,
$T$ fixed, and $X''=X(t''),X'=X(t'),x_r''=x_r(t''),x_r'=x_r(t')$.
The path integral in the center-of-mass coordinate is coupled to the relative
coordinate $x_r$ via $\dot x_r\dot X$, thus the $X$-path integration is that
of a free particle with a magnetic field, which is time dependent but not
$X$-path dependent. The classical solution for the center-of-mass coordinate $X$ 
is found to be
\begin{equation}
  X_{\Cl}(t)=\dfrac{X''-X'}{T}(t-t')-\dfrac{d}{2D}\int_{t'}^t\dot x_r(t)\dt\enspace,
\end{equation}
with corresponding classical action
\begin{equation}
 S_{\Cl}[X_{\Cl}]=\dfrac{d}{4D}\dfrac{(X''-X')^2}{T}
     -\dfrac{b^2}{4dD}\int_{t'}^{t''}\dot x_r^2(t)\dt\enspace.
\end{equation}
We can therefore separate off the center-of-mass path integration $X$ according 
to \cite{GRSh,GROGOb}
\begin{eqnarray}       & &\!\!\!\!\!\!\!\!
  K(X'',X',x_r'',x_r';T)
  =\sqrt{d\over 4D\pi\i\hbar T}\,
  \exp\Bigg(\ih\dfrac{d}{4D}\dfrac{(X''-X')^2}{T}\Bigg)
         \nonumber\\   & &\!\!\!\!\!\!\!\!\qquad\times
  \pathint{x_r}\exp\Bigg\{\ih\int_{t'}^{t''}\Bigg[\dfrac{1}{4d}\dot x_r^2
  -\lambda\Big(1-\alpha\,\e^{-\beta x_r}\Big)^2\Bigg]\dt\Bigg\}
                  \\   & & \!\!\!\!\!\!\!\! 
  =\int_{\bbbr}\dfrac{\d K}{2\pi}
  \exp\Bigg(\i K(X''-X')-\i\hbar K^2\dfrac{1-\mu M\kappa^2}{2M(1-2\mu\kappa)}T
  -\ih\lambda T\Bigg)
         \nonumber\\   & &\!\!\!\!\!\!\!\! \qquad\times
  \pathint{x_r}\exp\Bigg\{\ih\int_{t'}^{t''}\Bigg[
  \viert\bigg(\dfrac{1}{2\mu}-\kappa\bigg)^{-1}\dot x_r^2
  -\lambda\alpha^2\bigg(\e^{-2\beta x_r}-\dfrac2\alpha \e^{-\beta x_r}
  \bigg)\Bigg]\dt\Bigg\}\enspace.
         \nonumber\\   & &
\label{pathxr}
\end{eqnarray}
Again, the case $\kappa=0$ gives complete decoupling (note $1/4d=\mu/2$ for
$\kappa=0$).

The path integral in the second line of (\ref{pathxr}) is that of the Morse
potential which can be found in the literature \cite{GRSh}. Setting
\begin{eqnarray}
  K(X'',X',x_r'',x_r';T)&=&\int_{\bbbr}\dfrac{\d K}{2\pi}\,\e^{ \i K(X''-X')}
  K_K(x_r'',x_r';T)\enspace,
             \\  
  K_K(x_r'',x_r';T)&=&\int_{\bbbr}\dfrac{\d E}{2\pi\i}\,\e^{-\i ET/\hbar}
  G_K(x_r'',x_r';E)\enspace,
\end{eqnarray}
where $G_K$ is the (energy-dependent) Green function corresponding to the
kernel $K_K$. We find by means of the space-time transformation (Duru--Kleinert 
transformation) \cite{GRSh,DKa,KLEo} for the Green function
\begin{eqnarray}       & &\!\!\!\!\!\!\!\! \!\!\!\!\!\!\!\!
G_K(x_r'',x_r';E)=\ih\int_0^\infty\d T\,\e^{\i ET/\hbar}
  \exp\Bigg(-\i \hbar K^2\dfrac{1-\mu M\kappa^2}{2M(1-2\mu\kappa)}T
  -\ih \lambda T\Bigg)
         \nonumber\\   & &\!\!\!\!\!\!\!\! \!\!\!\!\!\!\!\!\qquad\times
  \pathint{x_r}\exp\Bigg\{\ih\int_{t'}^{t''}\Bigg[\dfrac{1}{4d}\dot x_r^2
  -\lambda\alpha^2\bigg(\e^{-2\beta x_r}-\dfrac2\alpha\,\e^{-\beta x_r}
  \bigg)\Bigg]\dt\Bigg\}\qquad
         \nonumber\\   & &\!\!\!\!\!\!\!\! \!\!\!\!\!\!\!\!
=\dfrac{\Gamma(\bhalf+\eta-\xi)}{2d\hbar^2\Gamma(1+2\eta)}\,
 \e^{-\beta(x_r'+x_r'')/2}\,W_{\xi,\eta}\Big(2\xi\,\e^{\beta x_{r,>}}\Big)
 M_{\xi,\eta}\Big(2\xi\,\e^{\beta x_{r,<}}\Big)\enspace.
\end{eqnarray}
Here I have abbreviated 
\begin{equation}
\xi=\dfrac1\hbar\sqrt{\lambda\over d}\enspace,\qquad
\eta=\dfrac1\hbar\sqrt{-\dfrac{E-E_0}{d}}\enspace,\qquad
E_0= \dfrac{\hbar^2K^2}{2M}\dfrac{1-\mu M\kappa}{1-2\mu\kappa}+\lambda\enspace,
\end{equation}
and $x_{r,\gtrless}$ denotes the larger/smaller of $x_r',x_r''$.
From the poles of $G_K$ the bound states are given by
\begin{equation}
  E_n=\dfrac{\hbar^2K^2}{2M}\dfrac{1-\mu M\kappa}{1-2\mu\kappa}
  -\bigg(\dfrac{1}{2\mu}-\kappa\bigg)\hbar^2\beta^2(n+\bhalf)^2
 +2\hbar\beta(n+\bhalf)\sqrt{\lambda\bigg(\dfrac{1}{2\mu}-\kappa\bigg)}\enspace.
\end{equation}
The first term is just the continuous spectrum which is not relevant for the
discrete spectrum.The bound state wave-functions are 
determined by the residua of the poles of $G_K$, and we find \cite{GRSh,DURa,GROb}
\begin{eqnarray}
  \Psi_{n,K}(X,x_r)&=&\dfrac{\e^{\i KX}}{\sqrt{2\pi}}
  \sqrt{\beta(2\xi-2n-1)n!(2\alpha\xi)^{2\xi-2n-1}\over\Gamma(2\xi-n)}
         \nonumber\\   &\ &\times
  \exp\Big((\xi-n-\bhalf)\beta x_r-\alpha\xi\,\e^{\beta x_r}\Big)
  L_n^{(2\xi-2n-1)}\Big(2\alpha\xi\,\e^{\beta x_r}\Big)\enspace.
\end{eqnarray}
The continuous spectrum is determined by the cut in $G_K$ and I obtain
\begin{equation}
  \Psi_{k,K}(X,x_r)=\dfrac{\e^{\i KX}}{\sqrt{2\pi}}
  \sqrt{\beta{k\sinh\pi k\over2\pi^2\alpha\xi}}\,
  \Gamma(\i k-\xi+\bhalf)\,\e^{-\beta x_r}
  W_{\xi,\i k}\Big(2\alpha\xi\,\e^{\beta x_r}\Big)\enspace,
\end{equation}
with the energy-spectrum
\begin{equation}
  E_k=\dfrac{\hbar^2K^2}{2M}\dfrac{1-\mu M\kappa}{1-2\mu\kappa}+
  \bigg(\dfrac{1}{2M}+\kappa\mu_2\mu_2\bigg)\hbar^2k^2\enspace.
\end{equation}

\section{The Double Pendulum}
As the next example we consider the double pendulum with two masses $m_1,m_2$
attached to two strings with lengths $l_1,l_2$. $g$ denotes the gravitational 
acceleration. Its Lagrangian is given by 
\begin{eqnarray}
  \CL&=&\dfrac{m_1+m_2}{2}\,l_1^2\dot\vphi_1^2+\dfrac{m_2}{2}
  \,l_2^2\dot\vphi_2^2
         \nonumber\\   & &\quad    
  +m_2l_1l_2\dot\vphi_1\dot\vphi_2\cos(\vphi_1-\vphi_2)
  +(m_1+m_2)gl_1\cos\vphi_1+m_2gl_2\cos\vphi_2\enspace.\qquad
\end{eqnarray}
In the harmonic approximation we have ($l_1=l_2=l$, $M=m_1+m_2$)
\begin{equation}
  \CL=\dfrac{M}{2}\,l^2\dot\vphi_1^2+\dfrac{m_2}{2}\,l^2\dot\vphi_2^2
  +m_2l^2\dot\vphi_1\dot\vphi_2
  -\dfrac{M}{2}\,gl\vphi_1^2-\dfrac{m_2}{2}\,gl\vphi_2^2\enspace,
  \end{equation}
which I consider in the sequel.
The path integral formulation for the latter case has the form
\begin{eqnarray}       & &\!\!\!\!\!\!\!\!
  K(\vphi_1'',\vphi_1',\vphi_2'',\vphi_2';T)
  =\pathint{\vphi_1}\pathint{\vphi_2}
         \nonumber\\   & &\!\!\!\!\!\!\!\!\qquad\times
  \exp\Bigg\{\ih\int_{t'}^{t''}\Bigg[
 \dfrac{M}{2}\,l^2\dot\vphi_1^2+\dfrac{m_2}{2}\,l^2\dot\vphi_2^2
  +m_2l^2\dot\vphi_1\dot\vphi_2
  -\dfrac{M}{2}\,gl\vphi_1^2-\dfrac{m_2}{2}\,gl\vphi_2^2
  \Bigg]\dt\Bigg\}\enspace.\qquad 
\label{kphi12}
\end{eqnarray}
The path integral (\ref{kphi12}) is of the form of a path integral for a general 
quadratic Lagrangian. Its solution is given by, e.g.~\cite{FH,GRSh,SCHUHd,KLEo,GROS}
\begin{equation}  
K(\vphi_1'',\vphi_1',\vphi_2'',\vphi_2';T)
=\bigg({1\over2\pi\i\hbar}\bigg)^{D/2}
  \sqrt{\det\bigg(-{\partial^2 S_{\Cl}(\vec\vphi'',\vec\vphi')\over
        \partial\vphi^{\prime\prime a}\partial\vphi^{\prime b}}\bigg)}
  \exp\bigg(\ih S_{\Cl}(\vec\vphi'',\vec\vphi')\bigg)\enspace,
\label{kphi12S}
\end{equation}
where I have denoted ${\vec\vphi}=(\vphi_1,\vphi_2)$, and 
$S_{\Cl}( {\vec\vphi}'', {\vec\vphi}')$ is the corresponding classical action 
of the system. The determinant appearing in (\ref{kphi12S}) is called
the Morette-Van Hove determinant \cite{GRSh,CHOQST}. In order to determine
the classical action, one considers the classical system of the coupled
differential equations and switches form the coordinates $(\vphi_1,\vphi_2)$
to normal coordinates $(\xi_1,\xi_2)$ by means of an affine transformation
\cite{LL}. This is done in the following way: We write
\begin{equation}
\CL=\half\dot{ \vec\vphi}\vec A \dot{ \vec\vphi}^t
 -\half{ \vec\vphi}\vec K{ \vec\vphi}^t\enspace,
\end{equation}
where $\vec A,\vec K$ are $n\times n$ matrices describing the classical system 
with its potentials and couplings. The classical system is diagonalized by 
considering the eigenvalue equation
\begin{equation}
  \det(\vec K-\omega^2\vec A)=0\enspace,
\end{equation}
with eigenvalues $\omega^2$. This is performed by a matrix $\vec C$, the affine
transformation. The corresponding differential equations in the normal coordinates 
$\vec\xi=\vec C\vec\vphi$ ($k=1,\dots,n$) are
\begin{equation}
  \ddot\xi_k+\omega^2_k\xi_k=0\enspace,
\end{equation}
with the $\omega_k$ the eigenfrequencies corresponding to the kth component.
In our case we have $n=2$, and we find
\begin{equation}
  \xi_1(t)=\sqrt{{M\over2}(1-r)}\,\Big(\vphi_1(t)-r\vphi_2(t)\Big)\enspace,\quad   
  \xi_2(t)=\sqrt{{M\over2}(1+r)}\,\Big(\vphi_1(t)+r\vphi_2(t)\Big)\enspace,
\label{xik}
\end{equation}
where $r=\sqrt{m_2/M}$, $\omega_{1,2}=(Mg/m_1l)(1\pm r)$. Vice versa we have
\begin{equation}
 \vphi_1(t)=\dfrac{1}{l\sqrt{2M}}\bigg(\dfrac{\xi_1(t)}{\sqrt{1-r}}
                           +\dfrac{\xi_2(t)}{\sqrt{1+r}}\bigg)  \enspace,\quad
 \vphi_2(t)=\dfrac{1}{l\sqrt{2M}}\bigg(\dfrac{-\xi_1(t)}{r\sqrt{1-r}}
                           +\dfrac{\xi_2(t)}{r\sqrt{1+r}}\bigg) \enspace.
 \end{equation}
Therefore we can write down the solution of the path integral (\ref{kphi12})
in the following form \cite{FEYb}--\cite{SCHUHd,KLEo}
\begin{eqnarray}       & & \!\!\!\!\!\!\!\! \!\!\!\!\!\!\!\!
K(\vphi_1'',\vphi_1',\vphi_2'',\vphi_2';T)
         \nonumber\\   & &\!\!\!\!\!\!\!\! \!\!\!\!\!\!\!\!
= \prod_{k=1,2}\sqrt{m_k\omega_k\over2\pi\i\hbar\sin\omega_kT}\,
 \exp\Bigg\{\ih\dfrac{m_k\omega_k}{2\sin\omega_kT}
   \Big[({\xi_k'}^2+ {\xi_k''}^2)\cos\omega_kT-2\xi'_k\xi''_k\Big]\Bigg\}\qquad\qquad
                  \\   & &\!\!\!\!\!\!\!\! \!\!\!\!\!\!\!\!
=\sqrt{m_1m_2\omega_1\omega_2\over(2\pi\i\hbar)^2\sin\omega_1T \sin\omega_2T}\,
         \nonumber\\   & &\!\!\!\!\!\!\!\! \!\!\!\!\!\!\!\!\qquad\times
\exp\Bigg\{\ih\dfrac{M}{4}\Bigg[\dfrac{m_1\omega_1}{\sin\omega_1T}(1-r)
    \Big((\vphi_1'-r\vphi_2')^2+ (\vphi_1''-r\vphi_2'')^2\Big)
         \nonumber\\   & &\!\!\!\!\!\!\!\! \!\!\!\!\!\!\!\! \qquad\qquad\qquad
    +\dfrac{m_2\omega_2}{\sin\omega_2T}(1+r)
    \Big((\vphi_1'+r\vphi_2')^2+ (\vphi_1''+r\vphi_2'')^2\Big)
         \nonumber\\   & &\!\!\!\!\!\!\!\! \!\!\!\!\!\!\!\!\qquad\qquad\qquad
    -\sqrt{1-r^2}\Big((\vphi_1'-r\vphi_2')(\vphi_1''-r\vphi_2'')
      + (\vphi_1'+r\vphi_2')(\vphi_1''+r\vphi_2'')\Big)\Bigg]\Bigg\}
                 \\   & &\!\!\!\!\!\!\!\! \!\!\!\!\!\!\!\!
=\prod_{k=1,2}\sum_{n_k=0}^\infty 
 \Psi_{n_k}^*(\vphi_k')\Psi_{n_k}(\vphi_k'')\e^{-\i(\omega_k+1/2)T}\enspace,
\label{wavefuncts}
\end{eqnarray}
The energy spectrum has the form
\begin{equation}
  E_{n_1,n_2}=\hbar(\omega_1n_1+\omega_2n_2+1)
     =\hbar\Bigg[\dfrac{Mg}{m_1l}\bigg(n_1+n_2+\sqrt{m_2\over M}\,(n_1-n_2)\bigg)+1
\Bigg]\enspace,
\end{equation}
and the wave-functions are given by ($\xi_k$ as in (\ref{xik}))
\begin{equation}
 \Psi_{n_k}(\vphi_k)=
   \bigg({m_k\omega_k\over\pi\hbar}\bigg)^{1/4}\sqrt{1\over2^{n_k}n_k!}
   \exp\bigg(-{m_k\omega_k\over2\hbar}\xi_k^2\bigg)
   H_{n_k}\bigg(\sqrt{m_k\omega_k\over\hbar}\,\xi_k\bigg)\enspace.
 \end{equation}
The expansion (\ref{wavefuncts}) can be derived by means of the Hille--Hardy
formula \cite{GRSh,GRA}.

\section{Conclusions}
The results clearly show the applicability of path integration techniques in
systems with kinetic energy couplings. The first example of a Morse potential
interaction, a model known in molecular physics, was transformed in center-of-mass
$X$ and relative coordinates $x_r$, followed by the path integration of the 
center-of-mass coordinate, thus achieving a complete separation of variables in 
$X$ and $x_r$. This left the remaining path integral as a path integral $K_K$ 
of the Morse potential, and the corresponding Green function $G_K$ could be written 
down. The usual analysis of $G_K$ gave the wave-functions and the energy spectrum 
of the discrete and continuous spectrum, respectively. The bound state energy 
spectrum has the typical feature of the energy spectrum of a Morse potential,
however modified by the coupling constant $\kappa$ of the $\kappa P_1P_2$-term.
Let us note that the choice of the Morse potential for the interaction in the relative
coordinate is but an example. We can take any potential $V(x_r)=V(|x_1-x_2|)$.
This has the consequence that in (\ref{pathxr}) the potential term
$\lambda\alpha^2\big(\e^{-2\beta x_r}-(2/\alpha)\e^{-\beta x_r}\big)$
is replaced by $V(x_r)$, and the corresponding path integral is exactly solvable for
any exactly solvable one-dimensional potential, modified by the mass-term $1/4d$.

The double pendulum in the harmonic approximation could be separated in terms of
normal coordinates $\vec\xi$ which are uniquely determined by the affine transformation 
$\vec C$ in terms of the original coordinates $\vec\vphi$. The path integral can be 
stated explicitly, and the wave-functions and the energy spectrum as well. The 
formulation in the normal coordinates is just the product of two one-dimensional 
harmonic oscillators, whereas the formulation in the original coordinates is somewhat 
more complicated, and features the interdependence of the two oscillations of the 
respective single pendula. Limiting cases, e.g.~$m_1\gg m_2$, are easily recovered 
yielding two (almost) independent oscillations ($m_2l^2g\dot\vphi_1\dot\vphi_2$ becomes
negligible). For the sake of simplicity I have considered only the case $l_1=l_2=l$. A 
generalization to $l_1\not=l_2$ is easily done, and left to the reader. The 
generalization  to higher dimensions is also straightforward; again we would obtain a 
product of $n$-dimensional harmonic oscillators in terms of normal coordinates, i.e., 
$\vec\xi=\vec C\vec q$, respectively $\vec q=\vec C^{-1}\vec\xi$, with $\vec C$ the
corresponding affine transformation.

These two examples thus show in a nice way the possibility of solving path
integrals for kinetic energy couplings. Other interesting cases would be
a particle in rotating magnetic fields, an arrangement important for trapping
electrons, and non-orthogonal coordinate systems in spaces of constant curvature
\cite{KALc}. This will be studied in subsequent investigations.



\bigskip\bigskip 
\vbox{\centerline{\ }
\centerline{\quad\epsfig{file=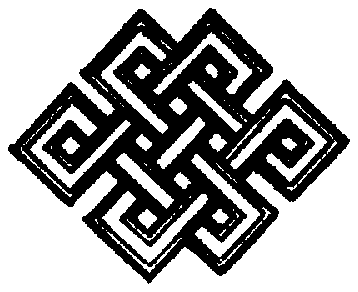,width=4cm,angle=90}}}

\end{document}